\documentclass[aps,manuscript]{revtex4}
\usepackage[T1]{fontenc}
\usepackage[latin1]{inputenc}
\usepackage{graphics}

\makeatletter

\providecommand{\LyX}{L\kern-.1667em\lower.25em\hbox{Y}\kern-.125emX\@}

\newcommand{\bee}{\begin{equation}}
\newcommand{\ee}{\end{equation}}
\newcommand{\beea}{\begin{eqnarray}}
\newcommand{\eea}{\end{eqnarray}}

\makeatother

\begin{document}

\noindent \hspace*{10cm} COLO-HEP-462\\
 \hspace*{10.5cm} April 2001.

\noindent \vspace*{1cm}

{\par\centering \textbf{\Large Flavor Symmetry and the Static Potential with
Hypercubic Blocking}\Large \par}

{\par\centering \vspace{0.5cm}\par}

{\par\centering {\large Anna Hasenfratz\( ^{\dagger } \) and Francesco Knechtli\( ^{\ddagger } \) }\large \par}

{\par\centering \vspace{0.5cm}\par}

{\par\centering Physics Department, University of Colorado, \\
 Boulder, CO 80309 USA\par}

{\par\centering \vspace{0.5cm}\par}

{\par\centering \textbf{Abstract}\par}

{\par\centering \vspace{0.5cm}\par}

{\par\centering We introduce a new smearing transformation, the hypercubic (HYP)
fat link. The hypercubic fat link mixes gauge links within hypercubes attached
to the original link only. Using quenched lattices at \( \beta =5.7 \) and
6.0 we show that HYP fat links improve flavor symmetry by an order of magnitude
relative to the thin link staggered action. The static potential measured on
HYP smeared lattices agrees with the thin link potential at distances \( r/a\geq 2 \)
and has greatly reduced statistical errors. These quenched results will be used
in forthcoming dynamical simulations of HYP staggered fermions.\par}

\vspace{0.5cm}


\vfill

\( ^{\dagger } \) {\small e-mail: anna@eotvos.colorado.edu}{\small \par}

\( ^{\ddagger } \) {\small e-mail: knechtli@pizero.colorado.edu}{\small \par}

\eject

\section{Introduction}

Fat link actions, actions where the thin link fermionic gauge connection is
replaced by a combination of extended but local gauge paths in a gauge invariant
manner, have been used in a variety of quenched and dynamical simulations. Since
the gauge paths of fat link actions combine only a finite number of gauge links
in the local environment of the original link, fat link actions are in the same
universality class as their thin link counterpart. For many physical observables
fat link actions show reduced scaling violations. Staggered fat link fermions
have improved flavor symmetry, both in quenched \cite{Blum:1997uf,Lagae:1998pe,Orginos:1999cr}
and dynamical simulations \cite{Orginos:1998ue,Karsch:2000ps,Knechtli:2000ku}.
Wilson-type clover fat link fermions have been studied only in quenched simulations
where they show improved chiral behavior resulting in better scaling of the
hadron spectrum and fewer exceptional configurations \cite{DeGrand:1998mn,Stephenson:1999ns}.
Fat links have been used in overlap simulations as a starting action for evaluating
Neuberger's formula \cite{Bietenholz:2000iy,DeGrand:2000tf,DeGrand:2000gq}.
The improved chiral properties of fat links result over an order of magnitude
improvement in convergence. 

It is quite intuitive to see why fat link staggered fermions show improved flavor
behavior. Flavor symmetry breaking of staggered fermions is the consequence
of distributing the components of the Dirac spinors to different lattice sites
within a hypercube. Since each lattice site couples to different gauge fields,
the different flavor and Dirac components feel different gauge environments.
The original \( SU(n_{f})\times SU(n_{f}) \) symmetry is explicitly broken
to a remnant \( U(1) \) symmetry. Unifying the gauge fields within the hypercube
would solve this problem. Smearing, while does not unify the gauge fields, can
remove some of the local fluctuations thus improving flavor symmetry. The most
effective smearing is one that maximally smoothes the gauge fields within the
hypercube but does not extend much beyond it.

The situation is similar for Wilson fermions that break chiral symmetry explicitly.
The Wilson Dirac operator is not anti-Hermitian, its eigenvalues are not constrained
to the imaginary axis. The real eigenvalues spread around \( \kappa _{cr} \)
causing spurious poles in the quenched quark propagator in the positive pion
mass region \cite{Bardeen:1998gv,Bardeen:1998eq,DeGrand:1998mn,Stephenson:1999ns}.
Since the real eigenvalue modes of the Dirac operator are instanton related,
one can study the spread of the real eigenvalues on smooth instanton backgrounds.
The real eigenmodes of the Wilson Dirac operator move farther from \( \kappa _{cr} \)
into the positive mass region on small instantons \cite{ DeGrand:1998mn}. The
Dirac operator has real modes even on configurations where the ``instanton'',
obtained by repeated renormalization group blocking from a large classical solution,
is small with respect to the lattice spacing. These objects do not have topological
interpretation, they are only dislocations, yet they are responsible for the
worst spurious poles of the Dirac propagator. On ordinary Monte Carlo gauge
configurations dislocations arise from large fluctuations of the plaquette.
Just as dislocations distort the topological charge measurements they also distort
the low lying Dirac spectrum and any physical observable related to it. Smearing
removes some of the fluctuations and their corresponding spurious eigenmodes.
This reduces the occurence of exceptional configurations, improves scaling \cite{DeGrand:1998mn,Stephenson:1999ns}
and makes topological charge measurement on pure gauge configurations possible
\cite{DeGrand:1998ss,Hasenfratz:1998qk}. Again, the most effective smearing
is one that maximally smoothes the gauge fields within a small, hypercube sized
volume, but does not extend much beyond it.

There are observables where some fat link actions have unacceptably large scaling
violations. Fattening changes the local, short distance properties of the gauge
configurations. On lattices with large lattice spacing too much fattening can
destroy short distance physical properties. This is most evident in the Coulomb
sector of the static potential \cite{DeGrand:1998ss}. Increased levels of smearing
destroy more and more of the short distance potential though it does not change
the long distance string tension. The heavy-light decay constants are also very
sensitive to smearing. Even low levels of smearing of the heavy fermions change
the value of \( f_{B} \) significantly \cite{Bernard:2000ki,Bernard:2000ht}.
Heavy fermions feel the short distance structure of the vacuum and that gets
distorted by the fattening. If only the light quark action is based on fat links,
this problem is avoided while the chiral properties of the light quarks are
improved. These expectations are supported by preliminary results \cite{Tomf_B}.
The fact that certain observables show large scaling violations with fat link
actions does not mean that fat link actions are incorrect, they are just not
effective in every situation.

It would be desirable to construct fat links that are local enough not to destroy
short distance quantities yet smooth enough to show the desired improved chiral
or flavor behavior of fat link actions. In this paper we describe a smearing
transformation, hypercubic blocking, along these lines. This transformation
improves flavor symmetry as much as three levels of APE blocking but mixes links
from hypercubes attached to the original link only. The static potential changes
only at small distances \( r/a<2 \) where lattice artifacts are anyhow large.
This paper deals with the definition and quenched properties of the hypercubic
fat link action. In a recent paper \cite{Knechtli:2000ku} we showed how to
simulate fat link dynamical systems. We were able to simplify and modify this
algorithm in such a way that dynamical simulations with the hypercubic action
are possible and efficient. We used the dynamical hypercubic action to study
finite temperature QCD with four flavors of staggered quarks \cite{Hyp_thermo}.

This paper is organized as follows. In Sect. 2 we define the hypercubic blocking.
The transformation has three tunable parameters. We describe how the parameters
are optimized and show that as the lattice spacing decreases the transformation
gets more effective, i.e. the number of smearing levels can be kept fixed and
small even as one approaches the continuum limit. We show that the hypercubic
action considerably improves the flavor symmetry of staggered quarks in quenched
simulations. The relative improvement with respect to the standard action is
about an order of magnitude and increases as the lattice spacing decreases.
In Sect. 3 we present quenched results for the static potential. We show that
the hypercubic action allows the determination of the Sommer scale \( r_{0} \)
and the string tension \( \sigma  \) with better accuracy compared to calculations
done with the thin link standard action. The distortion of the potential at
small distances due to the hypercubic blocking is limited to distances \( r/a<2 \)
independent of the lattice spacing and rotational symmetry is also improved.
In Sect. 4 we summarize our results.

\section{The Hypercubic Blocking}

\subsection{Definition and Optimization of the Hypercubic blocking}

\begin{figure}
{\par\centering \resizebox*{10cm}{!}{\includegraphics{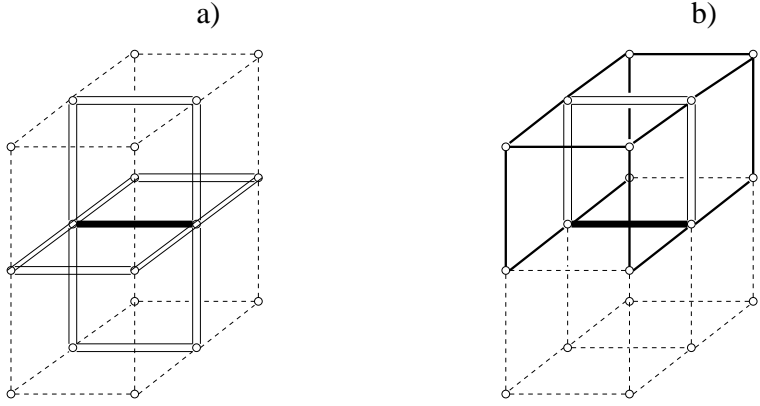}} \par}

\caption{Schematic representation of the hypercubic blocking in three dimensions. a)
The fat link is built from the four double-lined staples. b) Each of the double-lined
links is built from two staples which extend only in the hypercubes attached
to the original link. An important point is that the links entering the staples
are projected onto SU(3). \label{hypfig}}
\end{figure}
The fat links of the hypercubic blocking (HYP) are constructed in three steps.
At the final level the blocked link \( V_{i,\mu } \) is constructed via projected
APE blocking \cite{Albanese:1987ds} from a set of decorated links \( \tilde{V}_{i,\mu ;\nu } \)
as

\begin{equation}
\label{hyper1}
V_{i,\mu }=Proj_{SU(3)}[(1-\alpha _{1})U_{i,\mu }+\frac{\alpha _{1}}{6}\sum _{\pm \nu \neq \mu }\tilde{V}_{i,\nu ;\mu }\tilde{V}_{i+\hat{\nu },\mu ;\nu }\tilde{V}_{i+\hat{\mu },\nu ;\mu }^{\dagger }]\, ,
\end{equation}
 where \( U_{i,\mu } \) is the original thin link and the index \( \nu  \)
in \( \tilde{V}_{i,\mu ;\nu } \) indicates that the fat link at location \( i \)
and direction \( \mu  \) is not decorated with staples extending in direction
\( \nu . \) The decorated links \( \tilde{V}_{i,\mu ;\nu } \) are constructed
with a modified projected APE blocking from an other set of decorated links,
\( \bar{V}_{i,\mu ;\rho \, \nu } \) as 
\begin{equation}
\label{hyper2}
\tilde{V}_{i,\mu ;\nu }=Proj_{SU(3)}[(1-\alpha _{2})U_{i,\mu }+\frac{\alpha _{2}}{4}\sum _{\pm \rho \neq \nu ,\mu }\bar{V}_{i,\rho ;\nu \, \mu }\bar{V}_{i+\hat{\rho },\mu ;\rho \, \nu }\bar{V}_{i+\hat{\mu },\rho ;\nu \, \mu }^{\dagger }]\, ,
\end{equation}
 where the indices \( \rho \, \nu  \) indicate that the fat link \( \bar{V}_{i,\mu ;\rho \, \nu } \)
in direction \( \mu  \) is not decorated with staples extending in the \( \rho  \)
or \( \nu  \) directions. The decorated links \( \bar{V}_{i,\mu ;\rho \, \nu } \)
are constructed from the original thin links with a modified projected APE blocking
step
\begin{equation}
\label{hyper3}
\bar{V}_{i,\mu ;\nu \, \rho }=Proj_{SU(3)}[(1-\alpha _{3})U_{i,\mu }+\frac{\alpha _{3}}{2}\sum _{\pm \eta \neq \rho ,\nu ,\mu }U_{i,\eta }U_{i+\hat{\eta },\mu }U_{i+\hat{\mu },\eta }^{\dagger }]\, .
\end{equation}
 Here only the two staples orthogonal to \( \mu ,\, \nu  \) and \( \rho  \)
are used. With the construction eqs. (\ref{hyper1})-(\ref{hyper3}) the fat
link \( V_{i,\mu } \) mixes thin links only from hypercubes attached to the
original link. The hypercubic blocking is schematically represented in figure
\ref{hypfig}. The parameters \( \alpha _{1} \), \( \alpha _{2} \) and \( \alpha _{3} \)
can be optimized to achieve the smoothest blocked link configuration. The construction
eqs. (\ref{hyper1})-(\ref{hyper3}) can be iterated. 
\begin{figure}
{\par\centering \resizebox*{12cm}{!}{\includegraphics{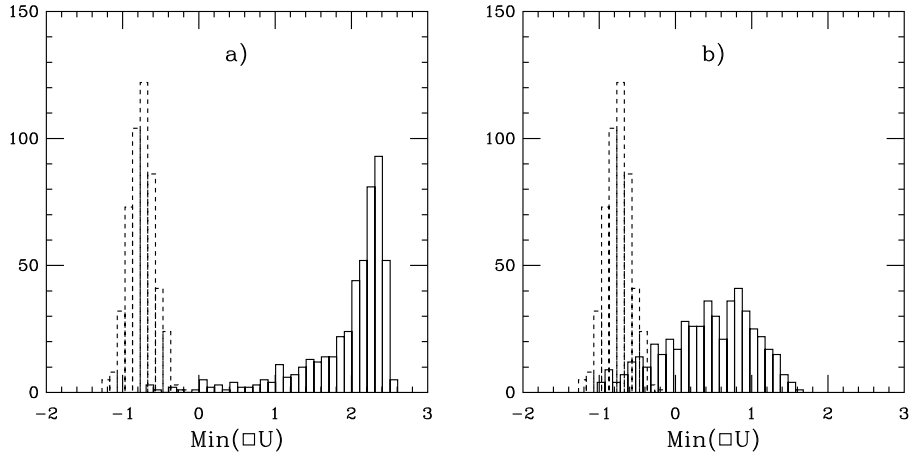}} \par}

\caption{\label{hist_b5.7_44}a) The histogram of the smallest plaquette distribution
on a set of \protect\( 4^{4}\protect \) \protect\( \beta =5.7\protect \) configurations
(dashed lines) and after hypercubic blocking (solid lines) on the same configurations.
b) Same as a) but the solid lines correspond to one level APE smearing with
\protect\( \alpha =0.7\protect \). }
\end{figure}

We could use a physical quantity like the level of flavor symmetry restoration
to optimize the parameters of the blocking, but in a three dimensional parameter
space that approach is very computer intensive. Instead we have decided to look
at the underlying reason of flavor and chiral symmetry violation. As we explained
in the introduction, both the staggered fermions' flavor symmetry violation
and Wilson fermions' exceptional configurations can be understood as fluctuations
in the local plaquette. It is not the average, not even the average fluctuation
of the plaquette that causes the problems but the few largest plaquette fluctuations
that create dislocations. To minimize the largest fluctuations we have to maximize
the smallest plaquettes of the system. We do this by considering the distribution
of the smallest plaquette in a finite, fixed volume lattice. The optimized hypercubic
blocking parameters will give the largest expectation value for the smallest
plaquette in this finite volume ensemble. We have optimized the \( \alpha  \)
parameters using a set of 500 \( 4^{4} \) lattices created at \( \beta =5.7 \)
with the Wilson plaquette action. 

Figure \ref{hist_b5.7_44}/a shows the histogram of the smallest plaquette on
each of the 500 original configurations (dashed lines) and after hypercubic
blocking (solid lines) with parameters
\begin{equation}
\label{alphaparams}
\alpha _{1}=0.75,\quad \alpha _{2}=0.6,\quad \alpha _{3}=0.3.
\end{equation}
The average of the smallest plaquette jumps from -0.75 to 1.95. This is about
the best that can be achieved with the hypercubic blocking. It is important
to remember that we are maximizing the smallest plaquette values on the lattice,
not the average plaquettes, which on these lattices are 1.67 and 2.82, respectively.
The minimum plaquette is not a well defined physical quantity, it strongly depends
on the lattice size, but as long as we compare different actions on the same
set of configurations, the minimum plaquette shows how the most severe vacuum
fluctuations change during smoothing. In figure \ref{hist_b5.7_44}/b we compare
the minimum plaquette distribution of the thin link action to a one-level APE
smeared action with \( \alpha =0.7 \) on the same configurations that were
used to obtain figure \ref{hist_b5.7_44}/a. Here the minimum plaquette average
increases only to 0.45 after smearing. Note that it is very important to project
to SU(3) the intermediate decorated links in eqs. (\ref{hyper1})-(\ref{hyper3}).
Without the projection the resulting hypercubic smearing is not better than
a one-level APE transformation.

\subsection{Flavor symmetry Restoration with Fat Link Actions}

\begin{figure}
{\par\centering \resizebox*{10cm}{!}{\includegraphics{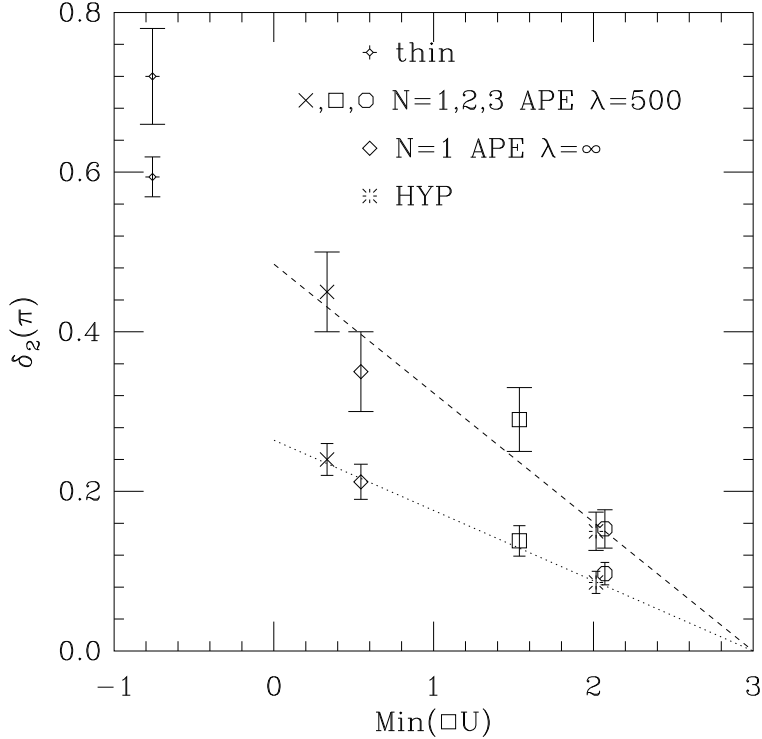}} \par}

\caption{\label{delta2 vs min plaq}The parameter \protect\( \delta _{2}\protect \)
of eq. (\ref{delta2}) for the two lightest non-Goldstone pions \protect\( \pi _{i,5}\protect \)
(lower points) and \protect\( \pi _{i,j}\protect \) (higher points) as the
function of the minimum plaquette average of the fat link actions. The pion
spectrum is computed on \protect\( 8^{3}\times 24\protect \), \protect\( \beta =5.7\protect \)
quenched configurations, the minimum plaquette average on \protect\( 4^{4}\protect \),
\protect\( \beta =5.7\protect \) configurations. The dotted (\protect\( \pi _{i,5}\protect \))
and dashed (\protect\( \pi _{i,j}\protect \)) lines show how the flavor symmetric
limit is approached as the minimum plaquette approaches three. }
\end{figure}
In the previous section we described an intuitive and fast way to optimize fat
link actions by maximizing the minimum plaquette average in finite volume lattices.
In this section we show that, as expected, this method optimizes flavor symmetry
as well. We compare the level of flavor symmetry restoration of a few fat link
actions with their respective minimum plaquette average computed on \( 4^{4} \)
configurations at \( \beta =5.7 \). We use the parameter \cite{Orginos:1998ue}
\begin{equation}
\label{delta2}
\delta _{2}=\frac{m^{2}_{\pi }-m^{2}_{G}}{m^{2}_{\rho }-m^{2}_{G}}
\end{equation}
 extrapolated to \( m_{G}/m_{\rho }=0.55 \) to quantify the difference between
the mass of the true Goldstone pion \( m_{G} \) and the mass of the pseudoscalar
particles or pions \( m_{\pi } \). For a flavor symmetric action \( \delta _{2}=0 \)
for all the pions \( \pi  \). We calculated the meson spectrum on an ensemble
of \( 8^{3}\times 24 \), \( \beta =5.7 \) quenched configurations using the
thin link, the N=1, 2, and 3 levels probabilistic APE blocked actions with \( \alpha =0.7 \)
and projection parameter \( \lambda =500 \) (for the definition of probabilistic
blocking and the projection parameter see Ref. \cite{Knechtli:2000ku}), the
N=1 deterministic (\( \lambda =\infty  \)) APE blocked action and the hypercubic
blocked action. Our results are summarized in figure \ref{delta2 vs min plaq}
where we plot \( \delta _{2} \) for the two lightest non-Goldstone pions, \( \pi _{i,5} \)
with flavor structure \( \gamma _{i}\gamma _{5} \) (lower points) and \( \pi _{i,j} \)
with flavor structure \( \gamma _{i}\gamma _{j} \) (higher points) as the function
of the minimum plaquette of the fat link actions. \( \delta _{2} \) seems to
depend on the minimum plaquette linearly approaching the flavor symmetric limit
\( \delta _{2}=0 \) as the dotted (\( \pi _{i,5} \)) and dashed (\( \pi _{i,j} \))
lines in figure \ref{delta2 vs min plaq} indicate, though we do not have a
theoretical explanation why this dependence should be linear. The HYP action
has about the same flavor symmetry violation than the N=3 APE action that we
studied in Ref. \cite{Knechtli:2000ku} reducing \( \delta _{2} \) by about
a factor of six relative to the thin link action.

\begin{figure}
{\par\centering \resizebox*{16cm}{!}{\includegraphics{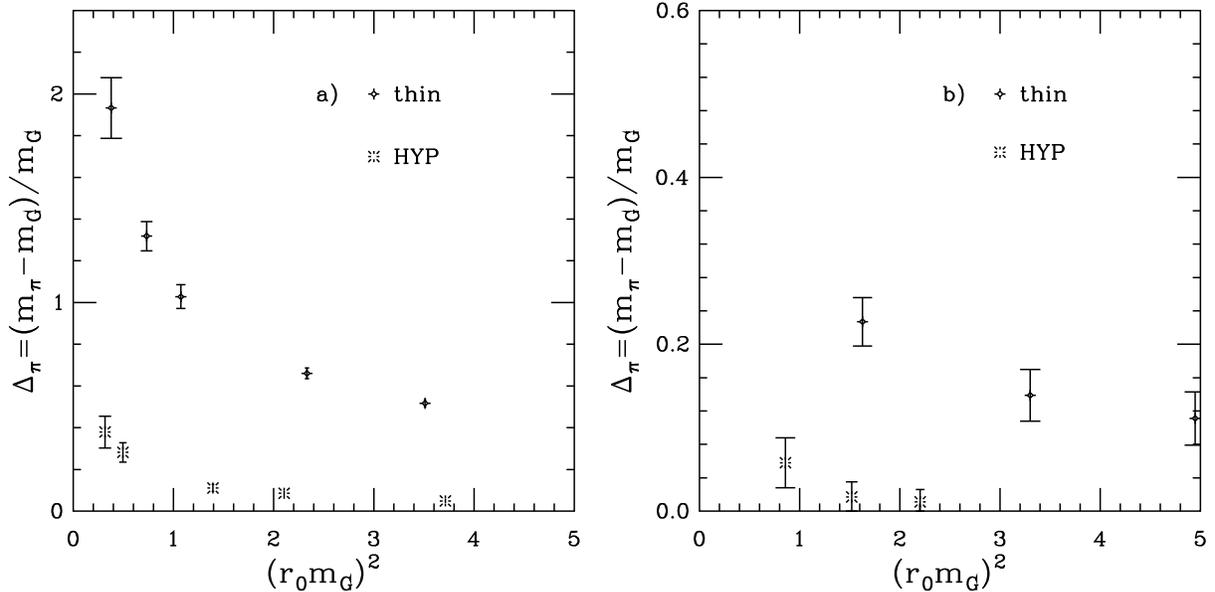}} \par}

\caption{\label{pisplit vs mq} a) The relative mass splitting \protect\( \Delta _{\pi }\protect \)
of eq. (\ref{Deltapi}) between the lightest non-Goldstone pion \protect\( \pi _{i,5}\protect \)
and the Goldstone pion \protect\( G\protect \) as the function of \protect\( (r_{0}m_{G})^{2}\protect \)
on \protect\( \beta =5.7\protect \) quenched configurations both for the thin
(fancy diamonds) and the HYP (bursts) actions. b) Same as a) but on \protect\( \beta =6.0\protect \)
quenched configurations where the lattice spacing is reduced by about a factor
of 2. The first three points at the lowest \protect\( r_{0}m_{G}\protect \)
values for the thin link action in both figures are from Ref. \cite{Gupta:1991mr}.}
\end{figure}
The data in figure \ref{delta2 vs min plaq} are extrapolated to \( m_{G}/m_{\rho }=0.55 \),
a value considerably higher than the physical value 0.175 of this ratio. Since
flavor symmetry violation increases as the the quark mass decreases, it is important
to see how far one can lower the quark mass before flavor symmetry is ruined.
In figure \ref{pisplit vs mq}/a we plot the relative mass splitting
\begin{equation}
\label{Deltapi}
\Delta _{\pi }=\frac{m_{\pi }-m_{G}}{m_{G}}
\end{equation}
 between the lightest non-Goldstone pion \( \pi _{i,5} \) and the Goldstone
pion \( G \) as the function of \( (r_{0}m_{G})^{2} \) on \( \beta =5.7 \)
quenched configurations both for the thin and HYP actions. The first three points
at the lowest \( r_{0}m_{G} \) values for the thin link action are from Ref.
\cite{Gupta:1991mr}. The ratio \( \Delta _{\pi } \) diverges as the quark
mass approaches zero but this divergence is much slower for the HYP action.
The relative mass splitting for the HYP action grows to about 30\% at \( r_{0}m_{G}=0.70 \)
where the ratio \( m_{G}/m_{\rho } \) of the Goldstone pion to the rho mass
is 0.23(8). At similar values of the \( m_{G}/m_{\rho } \) ratio the mass splitting
with the thin link action is between 130\% and 190\%.
\begin{figure}
{\par\centering \resizebox*{12cm}{!}{\includegraphics{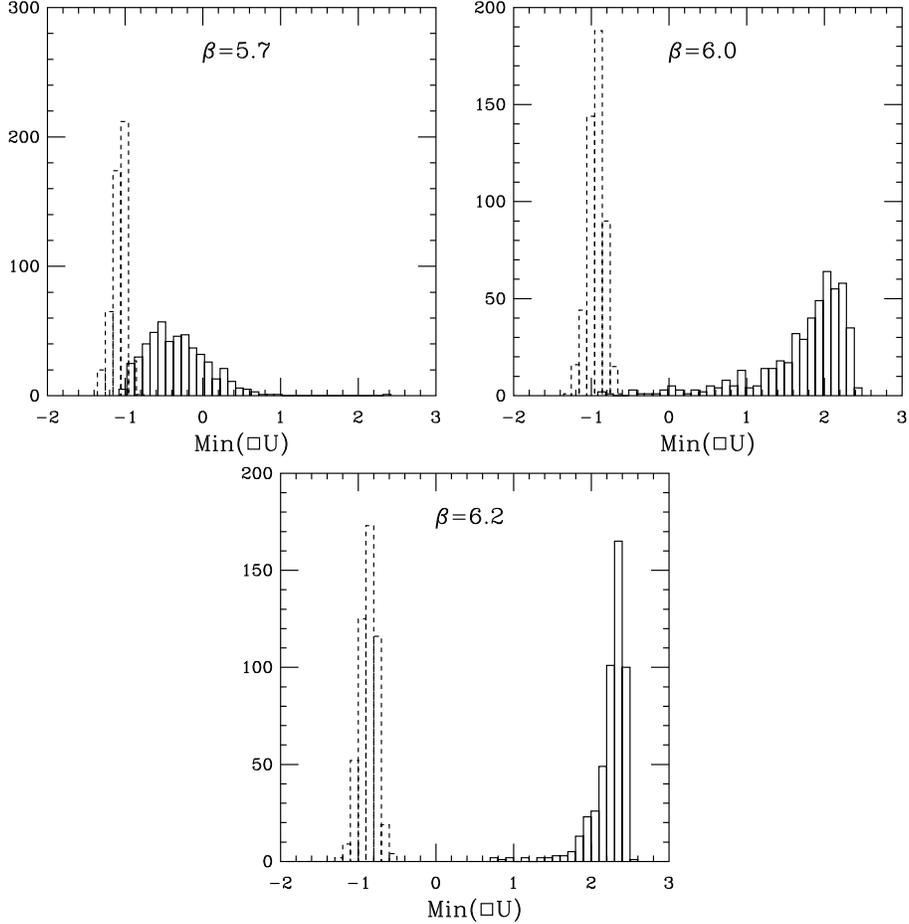}} \par}

\caption{The histograms of the smallest plaquette distribution on sets of \protect\( 8^{4}\protect \)
gauge configurations at different \protect\( \beta \protect \) values (dashed
lines) and after hypercubic blocking (solid lines) on the same configurations.
\label{hist_8888}}
\end{figure}

\subsection{Hypercubic blocking on Finer Lattices}

We have demonstrated that the minimum plaquette value strongly correlates with
the level of flavor symmetry restoration at the same lattice spacing. The question
naturally arises: if we decrease the lattice spacing, will the smearing be effective
or do we have to increase the number of smearing iterations? In figure \ref{hist_8888}
we show the minimum plaquette value calculated with the thin and hypercubic
actions with the optimized parameters of eqn. (\ref{alphaparams}) at \( \beta =5.7,6.0 \)
and 6.2. These calculations were done on \( 8^{4} \) lattices, that is why
the \textbf{\( \beta =5.7 \)} result looks significantly worse than before:
on a larger lattice the probability of finding a large plaquette fluctuation
is higher. The difference between the three \( \beta  \) values is striking.
One certainly does not need more smearing at a higher coupling. As \( \beta  \)
increases, the fat link minimum plaquette value increases faster than the corresponding
thin link minimum plaquette. That implies that one level of hypercubic blocking
will be effective for all coupling values. To verify this expectation we calculated
the HYP staggered meson spectrum on a set of \( 16^{3}\times 32 \), \( \beta =6.0 \)
configurations. The configurations were downloaded from the NERSC archive \cite{Kilcup:1997hp}.
The mass splitting \( \Delta _{\pi } \) for the lightest non-Goldstone pion
\( \pi _{i,5} \) is plotted as the function of \( (r_{0}m_{G})^{2} \) in figure
\ref{pisplit vs mq}/b. The thin link data are from Ref. \cite{Gupta:1991mr}.
The vertical scale of figure \ref{pisplit vs mq}/b is one forth of the vertical
scale of \ref{pisplit vs mq}/a but the qualitative features of the two plots
are very similar. \( \Delta _{\pi } \) decreases by an order of magnitude from
thin to HYP action, fairly independent of the lattice spacing or quark masses.

\section{The Static Potential of Fat link Actions}

In the following we consider the static potential, an observable that is quite
sensitive to the smearing of the gauge links. We analyzed two sets of lattices.
The first set consists 240 \( 8^{3}\times 24 \) configurations generated with
the Wilson plaquette action at \( \beta =5.7 \). The second set is an ensemble
of 220 \( 16^{3}\times 32 \) lattices from the NERSC archive generated also
with Wilson plaquette action at \textbf{\( \beta =6.0 \)} \cite{Kilcup:1997hp}\textbf{.}
We calculate the static quark potential on both sets using thin links and hypercubic
smeared links. 

The static potential on pure gauge lattices has been measured and analyzed by
several groups on larger lattices with orders of magnitude higher statistics.
Our goal here is not the improvement of existing potential measurements, rather
we want to demonstrate that the HYP action changes the potential only at small
distances, \( r/a<2 \) and, due to the smoothness of the gauge configuration,
provides an improved signal both for the Sommer parameter \( r_{0} \) \cite{Sommer:1994ce}
and even more for the string tension. The HYP potential also shows reduced rotational
symmetry violation. Our fitted values for \( r_{0}/a \) and \( a\sqrt{\sigma } \)
are consistent with previously published results. 

We extracted the potential by fitting the time dependence of the different Wilson
loops to a single exponential. We have measured the on-axis (m,0,0), \( m\leq n_{x}/2 \)
and three different off-axis sets: (m,m,0), \( m\leq n_{x}/2 \), (m,m,m), \( m\leq n_{x}/2 \)
and (m,2m,0), \( m\leq n_{x}/4 \) (\( n_{x} \) is the number of lattice points
in the spatial directions). To reduce statistical errors we used several (20-30)
levels of transversal APE smearing on the thin link lattices. After hypercubic
blocking the configurations are so smooth that transversal smearing is not important
anymore. We used only 5-10 levels of transversal smearing. The statistical errors
are from standard jackknife analysis. 

\begin{figure}
{\par\centering \resizebox*{16cm}{!}{\includegraphics{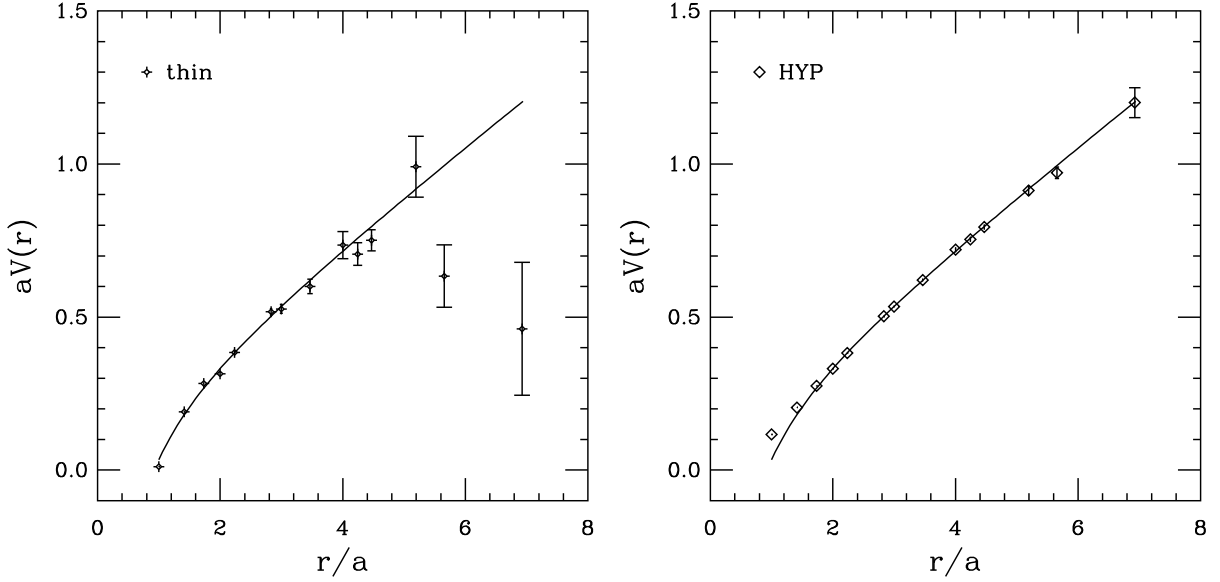}} \par}

\caption{\label{combined b=3D5.7}The thin link and HYP potentials for the \protect\( \beta =5.7\protect \)
data set. The thin link potential is shifted by a constant value to match the
HYP potential in the \protect\( 2.5<r/a<5.5\protect \) range. The solid line
is identical in both plots. It is a three parameter fit of the HYP potential
in the range \protect\( r/a\in (1.9,5.5)\protect \). }
\end{figure}
\begin{figure}
{\par\centering \resizebox*{16cm}{!}{\includegraphics{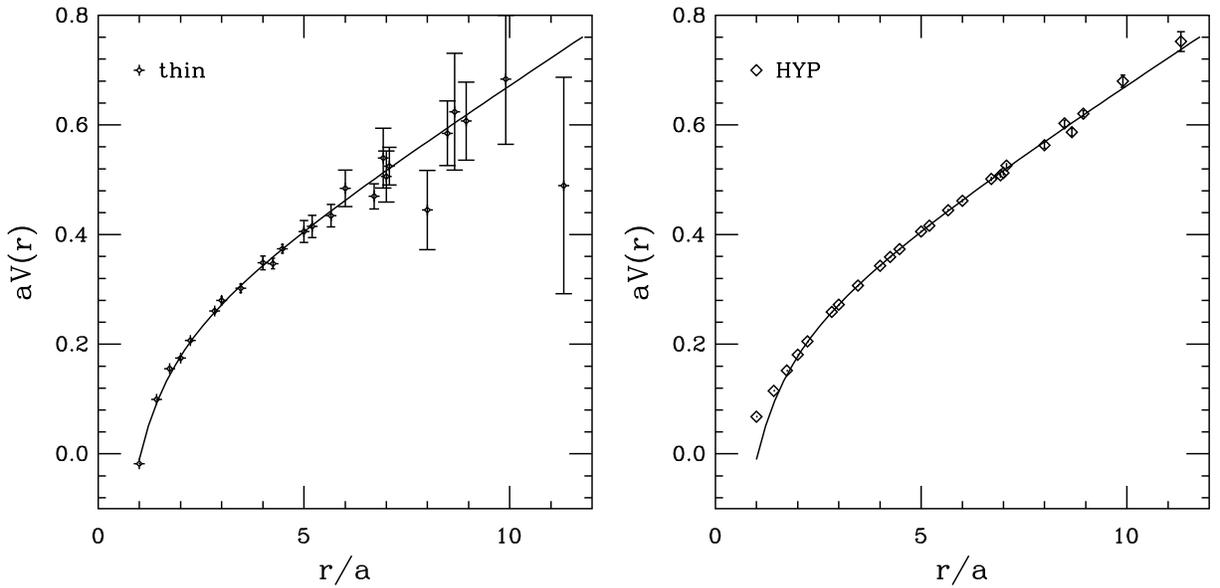}} \par}

\caption{\label{combined b=3D6.0}The thin link and HYP potentials for the \protect\( \beta =6.0\protect \)
data set. The thin link potential is shifted by a constant value to match the
HYP potential in the \protect\( 2.5<r/a<8\protect \) range. The solid line
is identical in both plots. It is a three parameter fit of the HYP potential
in the range \protect\( r/a\in (2.5,12)\protect \).}
\end{figure}

In figures \ref{combined b=3D5.7} and \ref{combined b=3D6.0} we plot both
the thin link and HYP potentials for the \( \beta =5.7 \) and \( \beta =6.0 \)
data sets. The thin link potentials are shifted by a constant value to match
the HYP potentials in the \( 2.5<r/a<5.5 \) (\( \beta =5.7 \) ) and \( 2.5<r/a<8 \)
(\( \beta =6.0 \)) range. Even though the first 5 \( r/a \) values were not
used for the matching in either coupling, the potentials match closely for \( r/a>\sqrt{2} \).
The distortion effects in the potential due to smearing are observable only
in the first two \( r/a \) values, independent of the lattice spacing. Moreover,
the HYP potentials have significantly reduced statistical errors, about a factor
of 4-7 at \( \beta =5.7 \), 10-15 at \( \beta =6.0 \). This is most obvious
at large distances. For example the \( \beta =6.0 \) thin link potential signal
becomes extremely noisy for \( r/a>7 \) with our statistics while the HYP potential
can be easily measured even at the largest distance, \( r/a=11.31 \) .
\begin{table}
{\centering \begin{tabular}{|c|c|c|c|c|}
\hline 
&
\( r_{0}/a \)&
\( a\sqrt{\sigma } \)&
\( N_{conf} \)&
\\
\hline 
\hline 
\( \beta =5.7 \)&
\( 2.96(2) \)&
\( 0.394(7) \)&
\( 240 \)&
HYP \\
\hline 
&
\( 2.990(24) \)&
\( 0.3879(39) \)&
\( 4000 \)&
Ref. \cite{Edwards:1998xf}\\
\hline 
&
\( 2.922(9) \)&
\( - \)&
\( 1000 \)&
Ref. \cite{Guagnelli:1998ud}\\
\hline 
\( \beta =6.0 \)&
\( 5.36(4) \)&
\( 0.218(2) \)&
\( 220 \)&
HYP \\
\hline 
&
\( 5.369(9) \)&
\( 0.2189(9) \)&
\( 4000 \)&
Ref. \cite{Edwards:1998xf}\\
\hline 
\end{tabular}\par}

\caption{\label{results}Results obtained with the HYP potential using a global fit
of the form of eq. (\ref{Eq. V(r)}). Also included are some large scale simulation
results at the same coupling values.}
\end{table}

To extract \( r_{0} \) and \( \sqrt{\sigma } \) from our data we fit the potential
in the form
\begin{equation}
\label{Eq. V(r)}
V(r)=\frac{c_{0}}{r}+c_{1}+c_{2}r
\end{equation}
 in a range \( r_{min}<r<r_{max}. \) \( r_{max} \) is limited by the accuracy
of our data, \( r_{max}/a=6-7 \) for the thin link, \( 10-12 \) for the HYP
potential at \( \beta =6.0 \). We choose \( r_{min} \) to exclude the region
where the potential is changed by the smearing or where short distance lattice
artifacts are large, typically \( r_{min}/a=1-3 \). We do not include any lattice
correction in the fit. For one, we do not know the perturbative potential for
the HYP action, though it can be calculated. On the other hand we do not seem
to need it as the HYP potential data shows only minimal lattice artifacts.
\begin{figure}
{\par\centering \resizebox*{10cm}{!}{\includegraphics{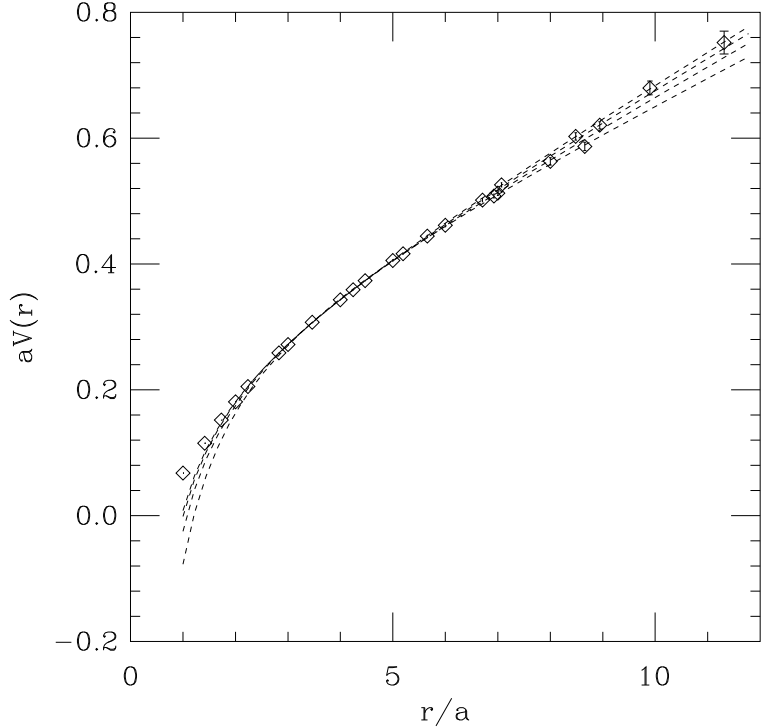}} \par}

\caption{\label{directional b=3D6.0}The HYP potential at \protect\( \beta =6.0\protect \).
The dashed lines are fits with eq. (\ref{Eq. V(r)}) along the different on-
and off-axis directions the potential is measured. All four fits use the range
\protect\( r/a\in (2.5,12)\protect \). }
\end{figure}
Our results obtained with the HYP potentials are summarized in Table \ref{results}.
Fits to our thin link potentials gave consistent results but with much higher
statistical errors and are not listed. In Table \ref{results} we also include
results from a few large scale potential simulations with thin links \cite{Guagnelli:1998ud,Edwards:1998xf}.
We considered several \( r_{max} \) and \( r_{min} \) values and included
all on- and off-axis data points in the given range in a global fit to eq. (\ref{Eq. V(r)}).
The errors in Table \ref{results} represent the average jackknife error of
the different fits. Our values for \( r_{0}/a \) and \( a\sqrt{\sigma } \)
from the HYP potential are consistent with the results from large scale computations.

We would like to emphasize that the HYP potential has about an order of magnitude
smaller statistical errors than the thin link potential on the same set of configurations.
This allowed us to extract the parameters of the potential with precision comparable
to the large scale simulations even though our analysis is not as refined. In
particular we did not use the one-link integral or multi-hit procedure \cite{Parisi:1983hm}
for the timelike links in the Wilson loops as has been done in \cite{Guagnelli:1998ud}.

In addition to the reduced statistical errors the HYP potential shows better
rotational invariance as well. We use the same quenched configurations both
for the thin and HYP potentials, but rotational symmetry violations can come
not only from the configurations but also from the operator used to measure
the potential. As an example we mention the potential for a perfect action \cite{DeGrand:1995ji}.
A perfect action has no scaling violations, yet the perturbatively calculated
potential is not rotational invariant unless one uses a perfect potential operator.
The HYP operator is of course not perfect, but it appears to be better than
the thin link Wilson loop operator. To illustrate rotational symmetry restoration,
we separated the on-axis and the three off-axis directions of the \( \beta =6.0 \)
potential and fitted them independently with the form of eq. (\ref{Eq. V(r)}).
The fits for the thin link potential were very poor but for the HYP potential
all four fits gave consistent results. We fitted the HYP potential in the range
\( r/a\in (2.5,12) \). The fitted values for \( r_{0}/a \) were all within
one standard deviation, for \( a\sqrt{\sigma } \) within two standard deviations.
The fit results are shown in figure \ref{directional b=3D6.0}. The four fitted
lines agree very well in the range \( r/a\in (2.5,6.5) \), they deviate at
larger \( r/a \) due to the slightly different \( a\sqrt{\sigma } \) values.
The uncertainty of the fitted lines is comparable to their spread in figure
\ref{directional b=3D6.0}. Closer inspection of the data points reveals some
rotational symmetry violation even with the HYP potential. It should be possible
to correct for some of these effects perturbatively thereby increasing the accuracy
of the scale determination.

The improved rotational symmetry of the HYP potential is not the main motivation
for using the HYP operator to measure the potential. Improved gauge actions
can be even more effective in restoring rotational invariance of the static
potential \cite{DeGrand:1995jk,Lepage:1996ph,Iwasaki:1997sn}. The improvement
of statistical accuracy by an order of magnitude of the HYP potential as compared
to the thin link potential is a more important feature.

\section{Summary}

In this paper we introduced a new kind of fat link, the hypercubic (HYP) fat
link. The HYP fat link is constructed to be localized, it does not extend beyond
hypercubes attached to the original link. The construction has three free parameters
that are optimized to reduce large plaquette fluctuations.

Using quenched configurations at \( \beta =5.7 \) and \( \beta =6.0 \) we
calculated the staggered spectrum of the HYP action. We showed that at both
coupling values flavor symmetry is greatly improved. The relative mass splitting,
\( \Delta _{\pi }=(m_{G}-m_{\pi })/m_{G} \) is reduced by an order of magnitude
at fixed \( r_{0}m_{G} \), independent of the coupling and the quark mass as
well. 

We have also considered the static potential calculated on HYP smeared \( \beta =5.7 \)
and \( \beta =6.0 \) configurations. We showed that smearing changes the potential
only at distances \( r/a<2 \) yet reduces statistical errors by about an order
of magnitude. Even with the very limited statistics of 220-240 lattices we could
extract both the Sommer parameter \( r_{0}/a \) and the string tension \( a\sqrt{\sigma } \).
Our results for both of these quantities are consistent with large scale simulation
results with errors that are better than our statistics would indicate from
the thin link potential.

We have undertaken this quenched study on the properties of the hypercubic blocking
because we want to do dynamical simulations with the HYP action. Such simulations
are feasible using combined over-relaxation and heat-bath updates. In a forthcoming
publication \cite{ Hyp_thermo} we use the dynamical HYP action in a finite
temperature simulation to show how improved flavor symmetry changes the phase
diagram of four-flavor dynamical staggered fermions.

\section{Acknowledgements}

We are grateful to Prof. T. DeGrand for many useful discussions during the course
of this work. This work was supported by the U.S. Department of Energy. We thank
the MILC collaboration for the use of their computer code.

\bibliographystyle{apsrev.bst}
\bibliography{hyp}

\end{document}